# Analytical Description of a Luminescent Solar Concentrator Device

## Ilya Sychugov*

*Department of Applied Physics, School of Engineering Sciences, KTH-Royal Institute of Technology, Stockholm, Sweden*
*Corresponding author: ilyas@kth.se*

**An analytical solution for the optical efficiency of a luminescent solar concentrator is presented. Due to a large number of input parameters and their complex effect on the device efficiency numerical simulations have been previously used for this purpose. The formulas, provided here, derived using a probabilistic approach, significantly reduce the complexity of the problem. The equations were validated by simulations and the obtained explicit expressions provide a clear common ground for the theoretical description of such devices. Implementation by a computer algebra program yields instant results for any set of input parameters. It allows a higher level of analysis, where an inverse task of finding parameters for a given efficiency can be readily solved.**

Radiation conversion by luminescence concentration has been considered decades ago, mainly motivated by a detector size reduction [1-3]. Acrylic glasses filled with organic dyes were proposed to convert incoming light to fluorescence for subsequent detection by small semiconductor photodetectors [4, 5]. With the development of detector technology and due to inherent limitations of the available fluorophores this method did not gain much traction. Recently, however, it has drawn renewed attention, stimulated by advances in colloidal quantum dot synthesis [6, 7].

The operation principle of a luminescent solar concentrator (LSC) is based on a total internal reflection of the re-emitted light for large angles, which is waveguided to the edges for conversion to electricity [8-13]. For this purpose a glass (plastic) slab is enriched with fluorophores, and photovoltaic cells are attached to the slab perimeter (Fig. 1). In evaluating the performance of such a device for, e.g. building-integrated photovoltaics, the power conversion efficiency is a key parameter. Multiple loss mechanisms, however, exist in the system. The re-emitted light can be absorbed by the matrix material, by the fluorophores, or it can be scattered out of the slab. The apparent complexity of the system stimulates the use of numerical simulations for the efficiency estimations [14-16].

Analytical treatment, if possible, can substantially reduce the complexity of the problem in the sense of a classical definition of complexity as the minimal length of a "code" needed to attain a result [17]. More importantly, explicit expressions can facilitate a unified benchmark tool, as opposite to undisclosed user-dependent codes. Finally, expedient results from the implementation by a computer algebra program would imply possibility of a higher level of analysis, such as solving an inverse problem of determining device parameters for a desired output efficiency.

In this paper an analytical solution has been derived by probabilistic treatment, including all the main loss mechanisms. It was based on the obtained distribution of the optical path lengths for an isotropic emitter randomly placed in a rectangular slab with absorbing edges. In this framework a result can be obtained instantly by a computer algebra program on a desktop computer as a continuous function of input parameters. Obtained values compare well with numerical simulations and represent a convenient approach for the complete analysis of LSC performance.

We start by considering an isotropic emitter randomly placed inside a rectangular slab with height $h$ and width $w$ (diagonal $d$) (Fig. 1). Isotropic emission corresponds to the light output pattern of quantum dots (QDs), which are typically spherical without specific dipole orientation [18]. For organic dyes this condition reflects a random orientation of the molecules uniformly distributed in the slab. It can be shown that in the plane of the slab (Fig. 1, left) the properly normalized probability density function (PDF) for a photon to travel an optical path $r$ is a piece-wise function (see section S1A in the Supplementary for the derivation):

$$p(r) = \begin{cases} \dfrac{2w + 2h - 2r}{\pi h w}, & 0 < r < w \\ \dfrac{2r - 2\sqrt{r^2 - w^2}}{\pi w r}, & w < r < h \\ \dfrac{2r^2 - 2h\sqrt{r^2 - w^2} - 2w\sqrt{r^2 - h^2}}{\pi h w r}, & h < r < d \end{cases} \quad (1)$$

This distribution is shown in Fig. 2 for two different geometries (blue and red lines). Points represent results of numerical simulations, where isotropic emitters were placed all over the slab and frequency counts for about a million of optical paths calculated. Numerical solutions indeed converge to the analytical formula $p(r)$.

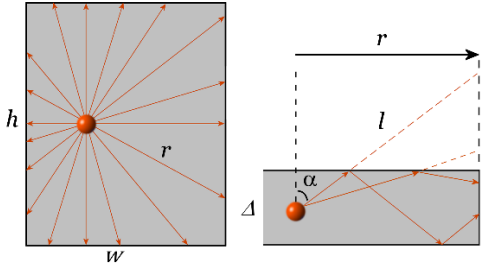

Fig. 1. Total optical path distribution from isotropic emitters in a rectangular slab (height $h$, width $w$, and thickness $\Delta$) stems from in-plane $p(r)$ and out-of-plane $q_r(l)$ distributions.

Next we note that together with the in-plane distribution there is a spread of optical paths in the plane perpendicular to the slab, as shown in Fig. 1, right (slab thickness $\Delta$). Here photons emitted along two directions are shown, both experiencing total internal reflection reaching the edge. Photons emitted to the escape cone (within the critical angle $\alpha_c$) directly contribute to the losses. For the typical refraction index of glass (polymers) $n = 1.5$ the critical angle is $\alpha_c \approx 42°$. From the diagram it is clear that the thickness of the slab plays no role in the out-of-plane optical path distribution ($\Delta \ll l$). Regardless of the first point of the total internal reflection (slab thickness) the optical path length (red dashed lines) depends only on the angle $\alpha$. For the out-of-plane distribution it can be shown that the distribution of optical path lengths for an isotropic emitter is (see section S1B in Supplementary for derivations):

$$q_r(l) = \frac{1}{2\pi} \cdot \frac{r}{l\sqrt{l^2 - r^2}}$$

for a given in-plane optical path length to the edge $r$. Convoluting both distributions one can obtain an expression for the full distribution of optical path lengths for the 3D case:

$$q(l) = \frac{1}{2\pi l} \int_{\frac{2}{3}l}^{l} \frac{p(r')r'}{\sqrt{l^2 - r'^2}} dr'$$

Integration limits come from the escape cone, where it can be shown that $r < l < nr = 3r/2$ (see section S1B in the Supplementary). This integral can be solved analytically using special functions (complete and incomplete elliptic integrals) and exact solutions are provided in the Supplementary section S1C. Those are presented graphically in Fig. 2, inset, as bright blue and red lines. It is seen that the distributions are smoothened and stretched to slightly longer values as compared to the 2D case. The solutions were again validated by simulations and numerical results do converge to the analytical expressions (Fig. S1). While the exact formulas are possible to obtain, the extensive presence of special functions make them not very practical to work with. To simplify the result to elementary functions one can note that the out-of-plane distribution $q_r(l)$ does not deviate far from the distance to the edge $r$ due to the limits set by the escape cone. So the average value $<l>$ can be taken instead of the distribution $q_r(l)$. It can be written as $l \approx <l> = k \cdot r$ (see supplementary S1B), where

$$k = \frac{2r(\ln(3 + \sqrt{5}) - \ln(2))}{\pi - 2\arctan(2/\sqrt{5})} \approx 1.14$$

Then the approximate solution for the 3D case can be represented simply through the 2D solution:

$$q(l) \approx p(l/k)$$

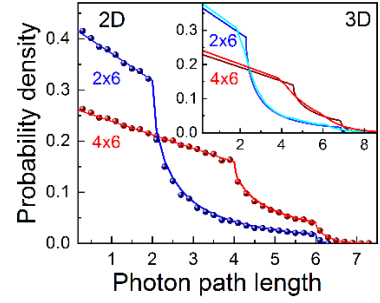

Fig.2 Distribution $p(r)$ of the optical path lengths for the photons from an isotropic emitter randomly placed in a 2D slab. Data points are simulated, and the lines are from Eq. (1) ($h = 6$ and $w = 4$ or $2$ length units). Inset shows full 3D distribution $q(l)$: bright lines are the exact solutions, and dark lines are the approximate solutions from Eq. (2).

A properly normalized PDF then can be written *explicitly* as:

$$q(l) = \begin{cases} \dfrac{2w + 2h - 2l/k}{\pi hwk}, & 0 < l < kw \\[6pt] \dfrac{2l - 2\sqrt{l^2 - (kw)^2}}{\pi lwk}, & kw < l < kh \\[6pt] \dfrac{2l^2/k - 2h\sqrt{l^2 - (kw)^2} - 2w\sqrt{l^2 - (kh)^2}}{\pi lhwk} & kh < l < kd \end{cases} \quad (2)$$

This function is shown in Fig. 2, inset, as dark blue and red lines. It reveals that the approximate solutions nearly coincide with the exact ones, indicating only a minor influence from the introduced assumption. The practical meaning of $q(l)$ is that the probability for a photon from an isotropic emitter randomly placed in the slab to experience an optical path $l$ would be $q(l)dl$.

From this expression one can already get some insight into the effect of the slab geometry. The probability of having an optical path below the slab width (aspect ratio $\beta = w/h$) can be calculated as $P_w \approx 0.31\beta + 0.56$ (see Supplementary, section S2B). It becomes clear that the rectangle width in fact limits most of the photon paths (Fig. S1, inset). For example, for the slab with a "golden ratio" $\beta \approx 0.62$, often used in architectural design of windows, $\sim 75\%$ of photons will travel distances shorter than the slab width.

With the optical path length distribution established the effect of matrix absorption on device efficiency can be readily evaluated. Let the linear absorption coefficient of the matrix be $\alpha\ [cm^{-1}]$. The probability of reaching the edge for a photon travelling distance $l'$ is $\exp(-\alpha l')$. It signifies the random nature of the absorption process with a given average rate $\alpha$ (experimental observation of this stochastic process is sometimes referred to as the Beer-Lambert law). Then the total probability of reaching the edge:

$$f(\alpha) = \int_0^{lmax} q(l') \cdot exp(-\alpha l')dl'$$

where $l_{max} = kd$. Calculating this integral yields (see Supplementary section S3 for derivations):

$$f(\alpha) = \frac{2}{\alpha^2 hw\pi k^2}((h+w)\alpha k - (kd\alpha + 1) \cdot e^{-kd\alpha} + e^{-kh\alpha} + e^{-kw\alpha} - 1) - \Im_1 - \Im_2 \quad (3)$$

where two integrals are:

$$\Im_1 = \frac{2}{w\pi k} \int_{kw}^{kd} \frac{\sqrt{l^2 - (kw)^2}}{l} exp(-\alpha l)\, dl, \quad (3a)$$

$$\Im_2 = \frac{2}{h\pi k} \int_{kh}^{kd} \frac{\sqrt{l^2 - (kh)^2}}{l} exp(-\alpha l)dl \quad (3b)$$

Formula (3) gives instant results for the effect of matrix absorption on the efficiency for a rectangular slab of any geometry and absorption coefficient, using a computer algebra program on a desktop computer (Fig. 3, left). One can compare with existing Monte-Carlo simulations using, e.g. results from [12] for PMMA ($\alpha_{PMMA} = 0.03$ cm$^{-1}$, blue dots) and for soda-lime glass ($\alpha_{WG} = 0.5$ cm$^{-1}$, red dots) in a square slab. Thus, analytical results of Eq. (3) coincide reasonably well with numerical simulations. The limitations of numerical Monte-Carlo method also become obvious: only one data point can be obtained per run and the proper convergence needs to be verified. The analytical result (3), on the other hand, produces a full functional dependence at once and changing geometry is just a matter of entering new values.

Separate from the matrix-induced losses the fluorophores themselves may attenuate re-emitted light. Consider first the effect of scattering. Let the linear scattering coefficient be $\alpha_{sc}$ [$cm^{-1}$]. It can be expressed through nanocrystal scattering cross-section $\sigma_{sc}$ and their concentration $N$ as $\alpha_{sc} = \sigma_{sc} N$. We invoke Rayleigh scattering on particles smaller than the wavelength, which is nearly isotropic. Probability of not being scattered after travelling distance $l'$ is $\exp(-\alpha_{sc} l')$. These photons will contribute to the total optical efficiency similarly to the absorption case above:

$$\chi_0(\alpha_{sc}) = f(\alpha_{sc})$$

In addition, there will be photons, which underwent a scattering event into the waveguiding mode. The probability of being scattered within a distance $l'$ is $1 - \exp(-\alpha_{sc} l')$. Let $\delta$ be the probability of scattering to the waveguided mode and not to the escape cone ($\delta \approx 75\%$ for $n = 1.5$, see Supplementary S4). Then the probability to reach the edge after one scattering event is (see Supplementary section S5 for derivations):

$$\chi_1(\alpha_{sc}) = \delta \cdot (1 - f(\alpha_{sc})) \cdot f(\alpha_{sc})$$

Summing up all contributions from multiple scattering events and using geometrical series one obtains:

$$\chi(\alpha_{sc}) = \sum_{i=0}^{\infty} \chi_i(\alpha_{sc}) = \frac{f(\alpha_{sc})}{1 - \delta \cdot (1 - f(\alpha_{sc}))}$$

One can quickly evaluate that for $\delta = 1$ (scattering without losses) the total probability is unity, as would be expected. For $\delta = 0$ (scattering with a complete loss) the expression becomes the same as for the absorption case. Also if the scattering coefficient $\alpha_{sc}$ is zero the optical efficiency is unity. An assumption of the Markov process was made here, so that the same optical path length distribution $q(l)$ could be used after every scattering event.

Similarly to the scattering loss the fluorophores can re-absorb propagating light with subsequent re-emission. To minimize this effect particles with a large Stoke shift are typically used in practice. Let the linear reabsorption coefficient be $\alpha_{re}$ [$cm^{-1}$]. Introducing a nanocrystal re-absorption cross-section $\sigma_{re}$ it can be represented as $\alpha_{re} = \sigma_{re} N$. If this process dominates the losses (no scattering or matrix absorption) it can be evaluated similar to the scattering as

$$\xi(\alpha_{re}) = \frac{f(\alpha_{re})}{1 - \delta \cdot QY \cdot (1 - f(\alpha_{re}))} \quad (4)$$

where $QY$ is the quantum yield ($QY \leq 1$). Here losses after every reabsorption event come not only from the escape cone, but also from the imperfect light conversion of emitters. An additional complication is a spectral dependence of the re-absorption coefficient $\alpha_{re} = \alpha_{re}(\lambda)$, due to a wavelength-dependent overlap of the luminophore emission and absorption bands.

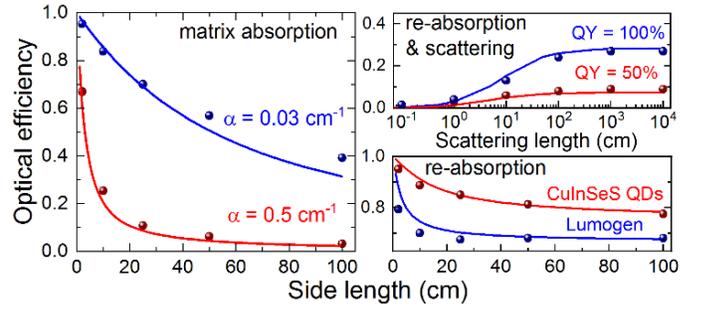

Fig. 3 Comparison of analytical solutions (Eq. (3) - (5)) for different loss mechanisms (lines) with numerical simulations (points) from [14, 15].

A spectral convolution can be used in this case with a properly normalized luminophore emission spectrum $I(\lambda)$:

$$\xi = \int_{\lambda min}^{\lambda max} I(\lambda) \cdot \xi(\alpha_{re}(\lambda)) d\lambda$$

As a first approximation, two re-absorption coefficients for the regions with strong and weak overlaps can be introduced. For example for Lumogen [14] a third of the emission band can be set experiencing $\alpha_{re1} = 1.5$ cm$^{-1}$, while the other two thirds are nearly reabsorption free: $\alpha_{re2} = 0$. The total optical efficiency in this case becomes a weighted sum $\xi(\alpha_{re1})/3 + 2/3$. Similarily for CuInSeS nanocrystals [14] a quarter of the band has $\alpha_{re3} = 0.3$ cm$^{-1}$ and for the rest $\alpha_{re4} = 0$, corresponding to the total efficiency $\xi(\alpha_{re3})/4 + 3/4$. In Fig. 3, right, bottom, analytical results for these luminophores are shown (red curve for CuInSeS QDs, and blue for Lumogen). It is seen that even without a proper spectral convolution the analytical results clearly reveal main features from the numerical Monte-Carlo simulations (dots).

So far we considered scenarios where a single loss mechanism dominates. In practice, they all can co-exist and their simultaneous contribution should be taken into account. Using similar combinatorics arguments as above (see Supplementary section S6 for derivations) it can be shown that for the case of scattering and matrix absorption co-existence the optical efficiency becomes:

$$\psi(\alpha_{sc}, \alpha) = \frac{f(\alpha_{sc} + \alpha)}{1 - \delta \cdot \frac{\alpha_{sc}}{\alpha_{sc} + \alpha} \cdot [1 - f(\alpha_{sc} + \alpha)]}$$

This formula turns into the expression for scattering only scenario for a non-absorbing matrix ($\alpha = 0$).

When re-absorption and scattering co-exist in the system:

$$\varphi(\alpha_{sc}, \alpha_{re}) = \frac{f(\alpha_{sc} + \alpha_{re})}{1 - \frac{\delta \cdot \alpha_{sc} + \delta \cdot QY \cdot \alpha_{re}}{\alpha_{sc} + \alpha_{re}} (1 - f(\alpha_{sc} + \alpha_{re}))} \quad (5)$$

To validate this derivation and result a comparison with numerical Monte-Carlo simulations from [15] is shown in Fig. 3, right, top. Here the optical efficiency as a function of the scattering length ($l_{sc} = 1/\alpha_{sc}$) is presented for Si QDs with a quantum yield of 50% (red) and 100% (blue) for a square slab 1x1 m² using $\alpha_{re} = 0.08$ cm$^{-1}$. For this type of fluorophores the wavelength dependence of the re-absorption within the emission band is small due to a very large Stoke shift [19, 20]. As in the numerical Monte-Carlo simulations [15], here the optical efficiency including the first absorption event is shown, i.e. $\delta \cdot QY \cdot \varphi(\alpha_{sc}, \alpha_{re})$. Again, the agreement with numerical Monte-Carlo simulations is reasonable.

Combining all loss mechanisms a general solution becomes

$$g(\alpha_{sc}, \alpha_{re}, \alpha) = \frac{f(\alpha_{sc} + \alpha_{re} + \alpha)}{1 - \frac{\delta \cdot \alpha_{sc} + \delta \cdot QY \cdot \alpha_{re}}{\alpha_{sc} + \alpha_{re} + \alpha}(1 - f(\alpha_{sc} + \alpha_{re} + \alpha))} \quad (6)$$

So equations (3) and (6) fully describe the effect of propagation losses in an LSC device. Input parameters are scattering ($\alpha_{sc}$), re-absorption ($\alpha_{re}$), and matrix absorption ($\alpha$) coefficients (first two can be expressed through luminophore concentration $N$ and cross-sections $\sigma_{sc}, \sigma_{re}$), geometry of the rectangular slab: height and width ($h, w$), fraction of the emission to the waveguiding mode $\delta$ ($\delta = 75\%$ for $n = 1.5$), quantum yield of the fluorophore ($QY$), and the correction factor for 3D geometry $k \approx 1.14$. These formulas can replace Monte-Carlo calculations and provide a quick and transparent tool for a thorough device analysis. Below we apply them to solve an inverse problem of finding acceptable device parameters for achieving a pre-set power output.

The optical power output of the device $\gamma$ [$W$], as collected at the edges, can be written as:

$$\gamma = \Phi \cdot h \cdot w \cdot (1 - T) \cdot \delta \cdot QY \cdot \eta \cdot g(\alpha_{sc}, \alpha_{re}, \alpha) \quad (7)$$

where $\Phi$ is the incoming energy flux [$W/cm^2$], $T$ is a transmitted fraction of the incoming sunlight ($1 - T$ is the absorbed fraction), and $\eta = \epsilon_{PL}/\epsilon_{sun}$ is the energy conversion coefficient of the luminescence. Product $\delta \cdot QY$ signifies losses after the first absorption event, i.e. before the waveguiding mode is initiated. Transmission of the visible light though the device is $T = \exp(-\alpha_{vis}\Delta) = \exp(-\sigma_{vis}N\Delta)$, where $\alpha_{vis}$ and $\sigma_{vis}$ are the fluorophore linear absorption coefficient and the absorption cross-section in the visible range, respectively (reflection neglected). This effectively gives 3 more input parameters: $\eta$, $\Delta$, and $\sigma_{vis}$. If necessary, spectral dependence of the re-absorption coefficient can be included by convolution with the normalized emission spectrum $I(\lambda)$. So considering $\delta, k$, and $\Phi$ as constants, in total there are 11 independent input parameters, where at least 3 are, in general, have wavelength dependence: $I(\lambda), \alpha_{re}(\lambda)$, and $\sigma_{vis}(\lambda)$.

We plotted Eq. (7) as a function of the aspect ratio $\beta = w/h$ for a slab with $h = 100\ cm$ and Si QDs as fluorophores (QY = 1), which are often considered for this application [15, 20, 21]. For these nanocrystals the photon energy conversion factor can be set $\eta \approx 0.6$ for the average solar photon energy of 2.5 eV and the fluorophore luminescence peak at 1.5 eV. Other parameters include the re-absorption coefficient $\alpha_{re} = 0.03\ cm^{-1}$ [20] and the scattering coefficient $\alpha_{sc} = 0.001\ cm^{-1}$ [15] for 0.1 wt. % concentration (5 nm diameter QDs). The solar energy flux is taken as $\Phi = 0.1\ W/cm^2$ and the device transmission $T$ is set to 50%. In the absence of all propagation-related losses, i.e. $g(\alpha_{sc}, \alpha_{re}, \alpha) = 1$, the optical power conversion efficiency equals $(1 - T)\delta\eta = 22.5\%$, shown as a dashed straight line. We can also set a minimum acceptable threshold for the power conversion to 7% (lower dashed line). This roughly corresponds to 5% (50 W/m²) electrical power output, taking into account conversion losses at the last stage. The grey area in between these lines then shows an acceptable working range.

Several optical power curves for different matrix absorption coefficients are shown in Fig. 4, left. An example file generating these curves for common computer algebra programs is provided in the Supplementary for reader's convenience. As expected, the deviation from the loss-free case is growing with increasing aspect ratio and stronger matrix absorption. A larger device area does not improve the output power much after a certain point, where losses from propagation start dominating. From such a plot one can graphically solve an inverse problem of finding input parameters for a given

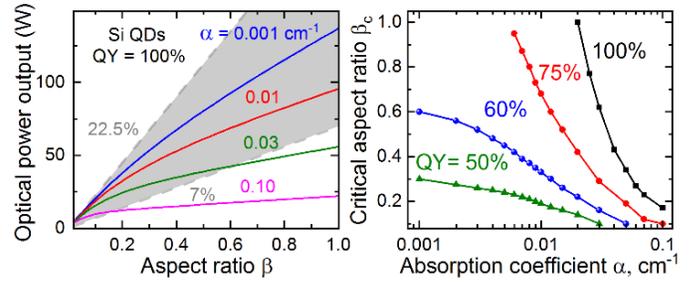

Fig. 4 (left) Device power output from the analytical solution of Eq. (7) for different parameters. (right) Critical aspect ratio $\beta_c$, resulting in 7% efficiency, for different quantum yields and absorption coefficients.

threshold efficiency (7% in this case). In Fig. 4, right, the critical aspect ratio $\beta_c$ corresponding to this efficiency is shown as a function of the matrix absorption coefficient for different quantum yield values. It is seen that for the "golden ratio" slab $\beta_c \approx 0.62$ and 60% quantum yield a very low matrix absorption coefficient is needed $\alpha = 0.001\ cm^{-1}$ (as in N-BK7 glass). Increasing quantum yield only by 15% relaxes this condition by an order of magnitude.

In conclusion, analytical formulas were derived to account for different losses in a luminescent solar concentrator device. The results were validated by numerical simulations of optical path distribution and propagation losses. The obtained solutions can be used to quickly and transparently evaluate LSC device performance for different material compositions and design [22-25] as well as for the description of light propagation in solar-pumped lasers [26].

**Funding.** Swedish Energy Agency (46360-1).

# Supplementary Information

# Analytical Solution of a Luminescent Solar Concentrator Device

Ilya Sychugov*

Department of Applied Physics, School of Engineering Science, KTH-Royal Institute of Technology, Stockholm, Sweden

* ilyas@kth.se

## S1. Derivation of the optical path length distribution in a slab

We are interested in the probability density for a photon to travel distance $r$ to the edge of a rectangular slab for an isotropic point-like source randomly placed inside it.

### A) 2D case in-plane (XY plane)

Consider edge element of a length $dy$. The fraction of the isotropic emission from a point source at distance $r$ into the edge element $dy$:

$$f = \frac{d\varphi}{2\pi}$$

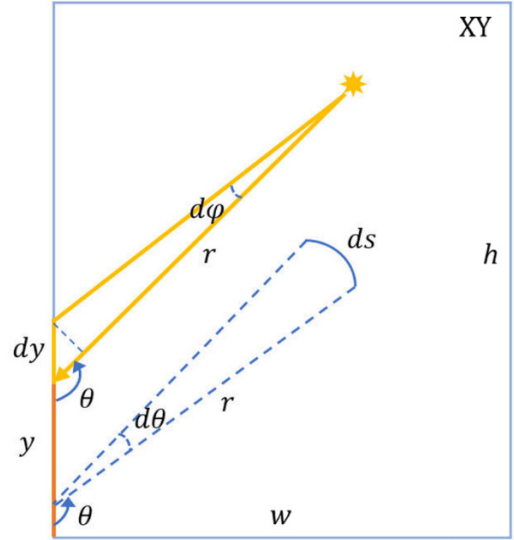

From the yellow triangle:

$$dy \cdot \sin(\pi - \theta) = 2 \cdot (r - dy \cdot \cos(\pi - \theta)) \cdot \sin(d\varphi/2)$$

For small $\varphi$ one can approximate $\sin(d\varphi/2) \approx d\varphi/2$.
For small $dy$ one can simplify $(r - dy \cdot \cos(\pi - \theta)) \approx r$.
Then
$$dy \cdot \sin(\theta) \approx r \cdot d\varphi$$

Then the fraction $f$ for a single source becomes:

$$f = \frac{dy \cdot \sin(\theta)}{2\pi r}$$

Elementary length $ds$ of the arc with radius $r$ contributing to the signal for the angle $\theta$ (blue segment):

$$ds = \frac{d\theta}{2\pi} \cdot 2\pi r = r d\theta$$

If the arc is limited by angles $\theta_{1,2}$ the total fraction of photons from the arc of a length $s$, arriving to the element $dy$ (summing up signal from all the sources at a distance $r$):

$$F = \int_0^s f ds = \frac{dy}{2\pi} \int_{\theta_1}^{\theta_2} \sin(\theta)\, d\theta = \frac{dy}{2\pi}(\cos(\theta_1) - \cos(\theta_2))$$

Finally, for the total fraction of photons reaching the edge (length $h$) after travelling distance $r$ one should integrate over the whole edge length:

$$p(r) = \frac{1}{2\pi} \int_0^h (\cos(\theta_1) - \cos(\theta_2))\, dy$$

which, after normalization, represents the probability density function. Since limiting angles $\theta_{1,2}$ vary depending on the geometry and the exact position of $dy$ several cases should be considered. For certainty a rectangular with a width smaller than the height ($w < h$) is taken into account.

**A1)** For $r < \frac{h}{2}$:

Three different cases can be considered:

1. For $0 < y < r$ the arc is from $\theta_1$ to $\pi$, where $\cos(\theta_1) = \frac{y}{r}$
2. For $r < y < h - r$ the arc is from $0$ to $\pi$
3. For $h - r < y < h$ the arc is from $0$ to $\theta_2$, where $\cos(\pi - \theta_2) = \frac{h-y}{r}, \cos(\theta_2) = -\frac{h-y}{r}$

And the total number of photons can be calculated by integrating respective parts:

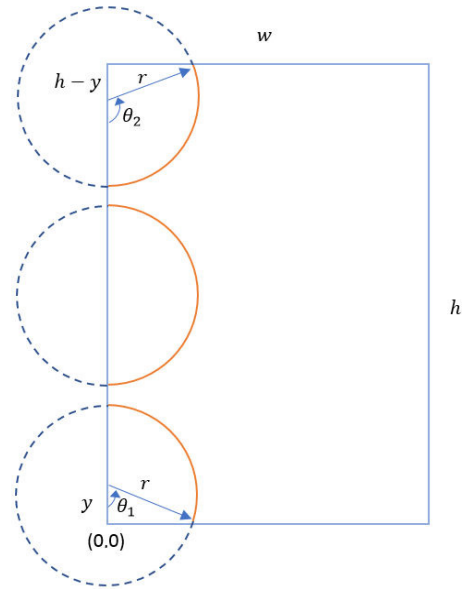

$$p_{left1}(r) = \frac{1}{2\pi}\left(\int_0^r \left(\frac{y}{r}+1\right)dy + \int_r^{h-r} 2\,dy + \int_{h-r}^h \left(1+\frac{h-y}{r}\right)dy\right) = \frac{2h-r}{2\pi}$$

**A2)** For $\frac{h}{2} < r < w$

Again three different cases can be considered:

1. For $0 < y < h - r$ the arc is from $\theta_1$ to $\pi$
2. For $h - r < y < r$ the arc is from $\theta_1$ to $\theta_2$
3. For $r < y < h$ the arc is from $0$ to $\theta_2$

$$p_{left2}(r) = \frac{1}{2\pi}\left(\int_0^{h-r}\left(\frac{y}{r}+1\right)dy \right.$$
$$+ \int_{h-r}^r \left(\frac{y}{r}+\frac{h-y}{r}\right)dy$$
$$\left.+ \int_r^h \left(1+\frac{h-y}{r}\right)dy\right)$$
$$= \frac{2h-r}{2\pi}$$

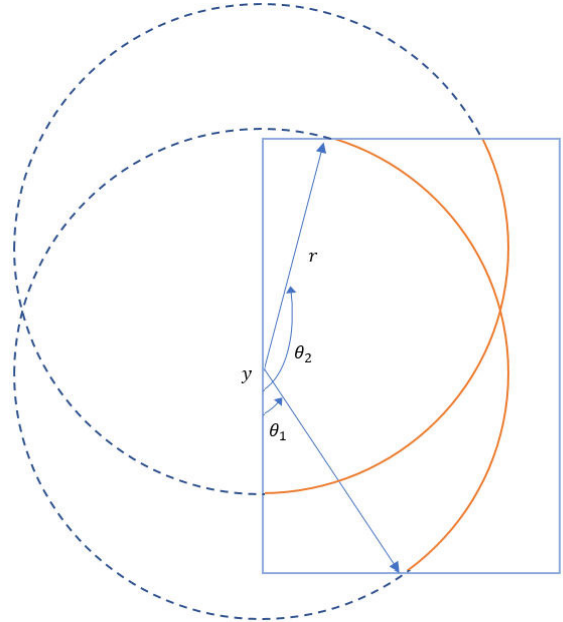

The same result as for the case above.

**A3)** For $w < r < h$

Two cases are:

1. $0 < y < h - r$ the arc is from 0 to $\theta_0$, where $\sin(\theta_0) = \frac{w}{r}, \theta_0 = \arcsin\left(\frac{w}{r}\right)$
2. $h - r < y < h - \sqrt{r^2 - w^2}$ the arc is from $\theta_3$ to $\theta_0$, where $\theta_3 = \pi - \theta_2 = \arccos\left(\frac{h-y}{r}\right)$

These two arcs repeat themselves from the other side, so their respective contributions should be multiplied by 2.

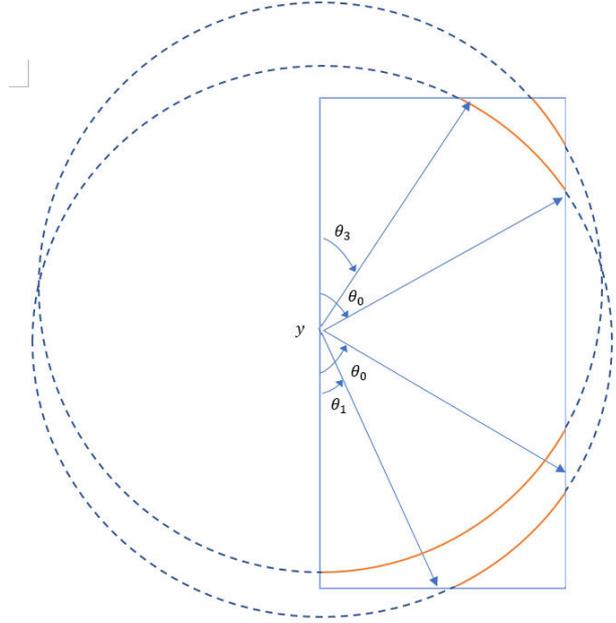

$$p_{left3}(r) = 2\int_0^{h-r}\left(1 - \cos(\arcsin\left(\frac{w}{r}\right))\right)dy + 2\int_{h-r}^{h-\sqrt{r^2-w^2}}\left(\frac{h-y}{r} - \cos(\arcsin\left(\frac{w}{r}\right))\right)dy$$

$$p_{left3}(r) = \frac{2hr - w^2 - 2h\sqrt{r^2 - w^2}}{2\pi r}$$

**A4)** For $h < r < \sqrt{h^2 + w^2}$

Two cases are:

1. For $0 < y < h - \sqrt{r^2 - w^2}$ the arc is from $\theta_3$ to $\theta_0$
2. For $\sqrt{r^2 - w^2} < y < h$ the arc is from $\theta_1$ to $\theta_0$

$$p_{left4}(r) = \int_0^{h-\sqrt{r^2-w^2}}\left(\frac{h-y}{r} - \cos(\arcsin\left(\frac{w}{r}\right))\right)dy + \int_{\sqrt{r^2-w^2}}^{h}\left(\frac{y}{r} - \cos(\arcsin\left(\frac{w}{r}\right))\right)dy$$

$$p_{left4}(r) = \frac{r^2 - w^2 - 2h\sqrt{r^2 - w^2} + h^2}{2\pi r}$$

One can repeat such derivations for the top edge.

**A5)** For $r < \frac{w}{2}$

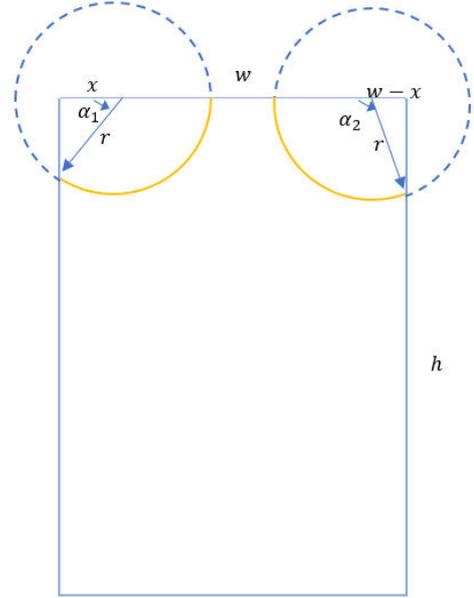

Similar to the first case for the left facet:
1. For $0 < x < r$ the arc is from $\alpha_1$ to $\pi$, where
$$\alpha_1 = \arccos\left(\frac{x}{r}\right)$$
2. For $r < x < h - r$ the arc is from $0$ to $\pi$
3. For $h - r < x < h$ the arc is from $0$ to $\alpha_2$, where
$$\alpha_2 = \arccos\left(-\frac{w-x}{r}\right)$$

$$p_{top1}(r) = \frac{2w - r}{2\pi}$$

**A6)** For $\frac{w}{2} < r < w$

1. For $0 < x < w - r$ countable arc is the same as above: from $\alpha_1$ to $\pi$
2. For $w - r < x < r$ it is from $\alpha_1$ to $\alpha_2$
3. For $r < x < w$ it is the same as above: from $0$ to $\alpha_2$

$$p_{top2}(r) = \frac{2w - r}{2\pi}$$

**A7)** For $w < r < h$ (same range as for left facet)

Only one case needs to be considered:
1. For $0 < x < w$ the countable arc is from $\alpha_1$ to $\alpha_2$

$$p_{top3}(r) = \int_0^w \left(\frac{x}{r} + \frac{w-x}{r}\right) dx = \frac{w^2}{2\pi r}$$

**A8)** For $h < r < \sqrt{h^2 + w^2}$

1. For $0 < x < w - \sqrt{r^2 - h^2}$ Countable arc is from $\alpha_3$ to $\alpha_0$, where $\alpha_0 = \arcsin\left(\frac{h}{r}\right)$, $\alpha_3 = \pi - \alpha_2 = \arccos\left(\frac{w-x}{r}\right)$
2. For $\sqrt{r^2 - h^2} < x < w$ Countable arc is from $\alpha_1$ to $\alpha_0$

$$p_{top4}(r) = \int_0^{w-\sqrt{r^2-h^2}} \left(\frac{w-x}{r} - \cos\left(\arcsin\left(\frac{h}{r}\right)\right)\right) dx + \int_{\sqrt{r^2-h^2}}^w \left(\frac{x}{r} - \cos\left(\arcsin\left(\frac{h}{r}\right)\right)\right) dx$$

$$p_{top4}(r) = \frac{r^2 - h^2 - 2w\sqrt{r^2 - h^2} + w^2}{2\pi r}$$

Then for the total perimeter of the rectangular (2 left and 2 top edges) a piecewise and continuous function $p(r)$ can be defined as (Figure 1):

$$p(r) = \begin{cases} p_1(r) = \dfrac{2w + 2h - 2r}{\pi}, & 0 < r < w \\ p_2(r) = \dfrac{2h(r - \sqrt{r^2 - w^2})}{\pi r}, & w < r < h \\ p_3(r) = \dfrac{2r^2 - 2h\sqrt{r^2 - w^2} - 2w\sqrt{r^2 - h^2}}{\pi r}, & h < r < \sqrt{h^2 + w^2} \end{cases}$$

Normalization coefficient is just the area of the rectangular $hw$, as would be expected:

$$\int_0^{\sqrt{h^2+w^2}} p(r)dr = \int_0^w p_1(r)dr + \int_w^h p_2(r)dr + \int_h^{\sqrt{h^2+w^2}} p_3(r)dr$$

$$\int_0^{\sqrt{h^2+w^2}} p(r)dr = \frac{w(w + 2h)}{\pi} + \frac{h}{\pi}\left(2h + (\pi - 2)w - 2\sqrt{h^2 - w^2} - 2w \cdot \arctan\left(\frac{w}{\sqrt{h^2 - w^2}}\right)\right) +$$

$$+ \frac{1}{\pi}\left(2h\sqrt{h^2 - w^2} + \pi wh - w^2 - 2h^2 - 2hw \cdot \arctan\left(\frac{w}{h}\right) - 2hw \cdot \arctan\left(\frac{h\sqrt{h^2 - w^2} - w^2}{w\sqrt{h^2 - w^2} + wh}\right)\right) =$$

$$= \frac{2hw}{\pi}\left(\pi - \arctan\left(\frac{w}{h}\right) - \arctan\left(\frac{w}{\sqrt{h^2 - w^2}}\right) - \arctan\left(\frac{h\sqrt{h^2 - w^2} - w^2}{w\sqrt{h^2 - w^2} + wh}\right)\right) = hw$$

Average value of the distribution:

$$\rho = <r> = \frac{\int_0^{\sqrt{h^2+w^2}} r \cdot p(r)dr}{\int_0^{\sqrt{h^2+w^2}} p(r)dr}$$

First moment (nominator):

$$\frac{1}{3\pi}\left(h^3 + w^3 - (h^2 + w^2)^{\frac{3}{2}} + 3hw^2 \ln\left(\frac{\sqrt{h^2 + w^2} + h}{w}\right) + 3wh^2 \ln\left(\frac{\sqrt{h^2 + w^2} + w}{h}\right)\right)$$

Then the average photon optical path from an isotropic emitter randomly placed in a rectangular:

$$\rho = \frac{h^3 + w^3 - (h^2 + w^2)^{\frac{3}{2}} + 3hw^2 \ln\left(\frac{\sqrt{h^2 + w^2} + h}{w}\right) + 3wh^2 \ln\left(\frac{\sqrt{h^2 + w^2} + w}{h}\right)}{3\pi hw}$$

For a simple case of a square slab ($h = w = a$):

$$\rho_{sq} = \frac{2(1 + 3\ln(1 + \sqrt{2}) - \sqrt{2})}{3\pi} a \approx 0.47a$$

## B) 2D case out-of-plane (XZ plane)

Emitted light from an isotropic emitter reflects many times from the media boundary due to the total internal reflection when the light is emitted outside the escape cone. Individual optical path length between reflections for the light emitted below critical angle $\theta_{c1}$ ( $\sin(\theta_{c1}) = 1/n$, $\theta_{c1} \approx 42°$ for $n = 1.5$ of glass or polymers):

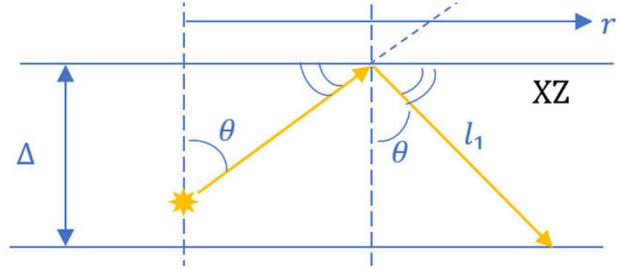

$$l_1 = \frac{\Delta}{\cos(\theta)}$$

Total optical path in this plane for $N$ bounces until reaching the edge

$$l = N \cdot l_1 = \frac{r}{l_1 \cdot \sin(\theta)} \cdot l_1 = \frac{r}{\sin(\theta)}$$

where $\theta_{c1} < \theta < \theta_{c2} = \pi - \theta_{c1}$. So $l$ does not deviate much from the distance to the edge $r$, and is in the range $r < l < nr = \frac{3}{2}r$ (for glass or polymers), depending on the angle $\theta$.

Probability density function for the light emitted from an isotropic emitter is constant $0 < \theta < 2\pi$:

$$g(\theta) = \frac{dP}{d\theta} = \frac{1}{2\pi}$$

Changing variables to the optical path length $l$ for a given parameter $r$

$$q_r(l) = \frac{dP}{dl} = \frac{dP}{d\theta} \cdot \left|\frac{d\theta}{dl}\right| = g(\theta) \cdot \left|\frac{d\theta}{dl}\right|$$

Where

$$\theta = \arcsin\left(\frac{r}{l}\right)$$

Therefore

$$\frac{d\theta}{dl} = -\frac{r}{l\sqrt{l^2 - r^2}}$$

Then we obtain

$$q_r(l) = \frac{1}{2\pi} \cdot \frac{r}{l\sqrt{l^2 - r^2}}$$

An average value of the optical path from the distribution $q_r(l)$ is close to $r$:

$$<l> = \frac{\int_r^{3r/2} q_r(l') \cdot l' dl'}{\int_r^{3r/2} q_r(l') dl'} = \frac{2r\left(\ln(3 + \sqrt{5}) - \ln(2)\right)}{\pi - 2\arctan(2/\sqrt{5})} \cdot r = k \cdot r \approx 1.144 \cdot r$$

## C) 3D case

The distance $r$ from the derived distribution above $q_r(l)$ is not a constant, but has a probability density distribution $p(r)$, where the probability of having $r = r'$ is $p(r')dr'$ for a properly normalized probability density function. That corresponds to the distribution of the optical path lengths:

$$q(l) = \frac{1}{2\pi l} \int_{\frac{2}{3}l}^{l} \frac{p(r')r'}{\sqrt{l^2 - r'^2}} dr'$$

Integration limits reflect the fact that only individual distributions with $\frac{2}{3}l < r' < l$ will contribute to the total probability density at the point $l$. Exact analytical solution is possible to obtain through special functions (complete and incomplete elliptic integrals). Using the following notations:

$$\eta = \frac{\sqrt{l^2 - w^2}}{l}, \varkappa = \sqrt{\frac{4l^2 - 9w^2}{4l^2 - 4w^2}}, \gamma = \frac{\sqrt{l^2 - h^2}}{l}, \chi = \frac{\sqrt{h^2 + w^2}}{l}$$

One can find that:

**For $0 < l < w$**

$$q_1(l) = \frac{1}{2\pi l} \int_{\frac{2}{3}l}^{l} \frac{p_1(r')r'}{\sqrt{l^2 - r'^2}} dr'$$

$$q_1(l) = \frac{(12h + 12w - 4l)\sqrt{5} - 9l\left(\pi - 2\arcsin\left(\frac{2}{3}\right)\right)}{36\pi^2} \approx 0.076(h + w) - 0.068 \cdot l$$

**If $w < \frac{2}{3}h$ then for $w < l < \frac{3}{2}w$ (otherwise for $w < l < h$)**

$$q_2(l) = \frac{1}{2\pi l} \int_{\frac{2}{3}l}^{w} \frac{p_1(r')r'}{\sqrt{l^2 - r'^2}} dr' + \frac{1}{2\pi l} \int_{w}^{l} \frac{p_2(r')r'}{\sqrt{l^2 - r'^2}} dr'$$

$$q_2(l) = \frac{1}{18\pi^2} \left( (6h + 6w - 2l)\sqrt{5} - 9l\left(\arcsin\left(\frac{w}{l}\right) - \arcsin\left(\frac{2}{3}\right)\right) - 9wk + 18h\left(\frac{w^2}{l^2}K(\eta) - E(\eta)\right) \right)$$

where $K, E$ are complete elliptic integrals of the first and second kind respectively.

**If $w < \frac{2}{3}h$ then for $\frac{3}{2}w < l < h$**

$$q_3(l) = \frac{1}{2\pi l} \int_{\frac{2}{3}l}^{l} \frac{p_2(r')r'}{\sqrt{l^2 - r'^2}} dr'$$

$$q_3(l) = \frac{1}{\pi^2} \left( \frac{h\sqrt{5}}{3} - hE(\eta) + h\frac{w^2}{l^2}\left(K(\eta) - F(\varkappa, \eta) + \Pi(\varkappa, \eta^2, \eta)\right) \right)$$

where $F, \Pi$ are incomplete elliptical integrals of the first and third kind respectively.

**If $w < \frac{2}{3}h$ then for $h < l < \sqrt{h^2 + w^2}$**

$$q_4(l) = \frac{1}{2\pi l} \int_{\frac{2}{3}l}^{h} \frac{p_2(r')r'}{\sqrt{l^2 - r'^2}} dr' + \frac{1}{2\pi l} \int_{h}^{l} \frac{p_3(r')r'}{\sqrt{l^2 - r'^2}} dr'$$

$$q_4(l) = \frac{1}{\pi^2} \left( \frac{h\sqrt{5}}{3} - \frac{h}{2l}\sqrt{l^2 - h^2} + \frac{\pi l}{4} + h\frac{w^2}{l^2}\big(\Pi(\varkappa, \eta^2, \eta) - F(\varkappa, \eta) + K(\eta)\big) - hE(\eta) - wE(\gamma) \right.$$

$$\left. - \frac{l}{2}\arcsin\left(\frac{h}{l}\right) + h^2\frac{w}{l^2}K(\gamma) \right)$$

**If $w < \frac{2}{3}h$ then for $\sqrt{h^2 + w^2} < l < \frac{3}{2}h$**

$$q_5(l) = \frac{1}{2\pi l} \int_{\frac{2}{3}l}^{h} \frac{p_2(r')r'}{\sqrt{l^2 - r'^2}} dr' + \frac{1}{2\pi l} \int_{h}^{\sqrt{h^2 + w^2}} \frac{p_3(r')r'}{\sqrt{l^2 - r'^2}} dr'$$

$$q_5(l) = \frac{1}{\pi^2} \left( \arcsin(\chi)\frac{l}{2} - \arcsin\left(\frac{h}{l}\right)\frac{l}{2} + h\frac{w^2}{l^2}\left(F\left(\frac{h}{\eta \chi l}, \eta\right) - \Pi\left(\frac{h}{\eta \chi l}, \eta^2, \eta\right) - F(\varkappa, \eta) + \Pi(\varkappa, \eta^2, \eta)\right) \right.$$

$$\left. + h^2\frac{w}{l^2}\left(F\left(\frac{w}{\eta \chi l}, \gamma\right) - \Pi\left(\frac{w}{\eta \chi l}, \gamma^2, \gamma\right)\right) - \frac{\chi}{2}\sqrt{l^2 - h^2 - w^2} - \frac{h}{2l}\sqrt{l^2 - h^2} \right)$$

To verify these formulas several millions of path lengths were numerically calculated for a point with a varying location inside a 3D slab with given side lengths. Resulting distributions (dots) indeed converge to the analytical expressions presented here (blue and red lines).

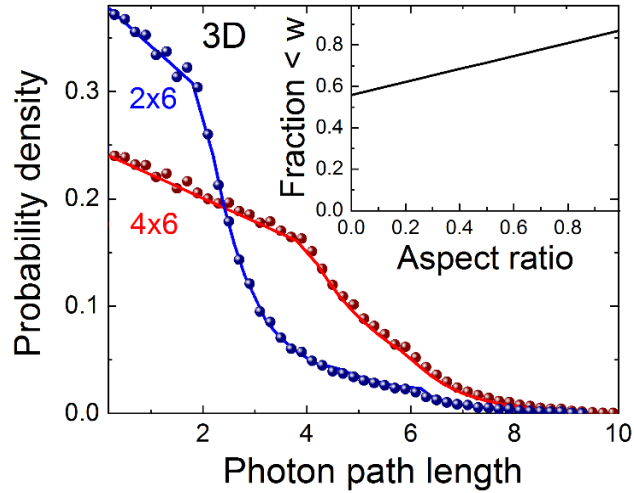

**Figure S1.** Probability density function distributions for a 3D slab with dimension 2x6 (blue) and 4x6 (red) units. Points are counted by simulating about a million paths from an isotropic emitter, and solid lines are analytical solutions from above. Inset shows fraction of optical paths below the width $w$ of a rectangular (aspect ratio $\beta = w/h$).

## S2. Approximate solution for the optical path distribution

### A) Derivation of the approximate solution

The exact solution presented above is not very convenient to work with, so an approximate analytical solution would be easier to use instead. It can be obtained based on the fact that $q_r(l)$ varies only marginally, being chiefly close to $r$. So, as a first approximation, one can substitute distribution $q_r(l)$ by its average value $l \approx <l> = k \cdot r$ and to rely solely on the obtained 2D distribution $p(r)$. So the approximate analytical distribution for the optical path length distribution in 3D can be written as

$$q(l) \approx p(l/k)$$

It appears to be a very good approximation for different aspect ratio geometries (Figure 1, inset). The meaning of the coefficient $k \approx 1.14$ can be then interpreted as a correction for 3D geometry from a 2D case. So the final solution becomes:

$$q'(l) = \begin{cases} \dfrac{2w + 2h - 2l/k}{\pi}, & 0 < l < kw \\ \dfrac{2h(l - \sqrt{l^2 - (kw)^2})}{\pi l}, & kw < l < kh \\ \dfrac{2l^2/k - 2h\sqrt{l^2 - (kw)^2} - 2w\sqrt{l^2 - (kh)^2}}{\pi l}, & kh < l < k\sqrt{h^2 + w^2} \end{cases}$$

Using normalization coefficient:

$$\int_0^{k\sqrt{h^2+w^2}} q'(l) dl = k \cdot \int_0^{\sqrt{h^2+w^2}} p(r) dr = khw$$

A properly normalized 3D probability density function $q(l)$ then becomes:

$$q(l) = \begin{cases} \dfrac{2w + 2h - 2l/k}{\pi hwk}, & 0 < l < kw \\ \dfrac{2l - 2\sqrt{l^2 - (kw)^2}}{\pi lwk}, & kw < l < kh \\ \dfrac{2l^2/k - 2h\sqrt{l^2 - (kw)^2} - 2w\sqrt{l^2 - (kh)^2}}{\pi lhwk}, & kh < l < k\sqrt{h^2 + w^2} \end{cases}$$

### B) Probability of the optical path to be shorter than the rectangular width

The probability for an optical path to be shorter than the rectangular width $w$ (aspect ratio $\beta = \frac{w}{h} \leq 1$):

$$P_w = \int_0^w q(l) dl = \frac{2}{k\pi}\left(\beta + 1 - \frac{\beta}{2k}\right) \approx 0.31\beta + 0.56$$

It is shown in the inset of Figure S1 as a function of $\beta$. So most of the photon path distribution lies below the shortest side of the rectangular. Even for a very large 1:5 ratio it is > 60% probability, reaching ~ 85% for the squared shape. For the "golden ratio" $\frac{2}{1+\sqrt{5}} \approx 0.62$ it is 75%. So for most practical applications it is possible to say that the rectangular width mainly limits optical path of photons in a 3D slab.

## S3. Effect of matrix absorption

The probability of having optical path $l = l'$ is $q(l')dl'$ for a properly normalized probability density function $q(l)$. Then

$$f(\alpha) = \int_0^{lmax} q(l') \cdot \exp(-\alpha l')dl'$$

Which essentially shows the fraction of photons reaching the edge for given $h$ and $w$ (diagonal $d = \sqrt{h^2 + w^2}$) of a rectangular, where $l_{max} = kd$. Calculating dimensionless $f(\alpha)$ using obtain normalized distribution $q(l)$ yields:

$$f_1(\alpha) = \int_0^{kw} q_1(l') \cdot \exp(-\alpha l')dl' = \frac{2}{\alpha^2 hw\pi k^2}\left((h+w)\alpha k - 1 + e^{-kw\alpha}(1 - kh\alpha)\right)$$

$$f_2(\alpha) = \int_{kw}^{kh} q_2(l') \cdot \exp(-\alpha l')\,dl' = \frac{2}{\alpha w\pi k}\left(e^{-kw\alpha} - e^{-kh\alpha}\right) - \frac{2}{w\pi k}\int_{kw}^{kh} \frac{\sqrt{l^2 - (kw)^2}}{l}\exp(-\alpha l)dl$$

$$f_3(\alpha) = \int_{kh}^{kd} q_3(l') \cdot \exp(-\alpha l')\,dl' = \frac{2}{\alpha^2 hw\pi k^2}(\alpha hk \cdot e^{-\alpha hk} - d\alpha k \cdot e^{-\alpha dk} + e^{-\alpha hk} - e^{-\alpha dk})$$
$$- \frac{2}{w\pi k}\int_{kh}^{kd} \frac{\sqrt{l^2 - (kw)^2}}{l}\exp(-\alpha l)\,dl - \frac{2}{h\pi k}\int_{kh}^{kd} \frac{\sqrt{l^2 - (kh)^2}}{l}\exp(-\alpha l)dl$$

$$f(\alpha) = f_1(\alpha) + f_2(\alpha) + f_3(\alpha)$$

$$f(\alpha) = \frac{2}{\alpha^2 hw\pi k^2}\left((h+w)\alpha k - (kd\alpha + 1) \cdot e^{-kd\alpha} + e^{-kh\alpha} + e^{-kw\alpha} - 1\right) - \mathfrak{I}_1 - \mathfrak{I}_2$$

where two integrals are:

$$\mathfrak{I}_1 = \frac{2}{w\pi k}\int_{kw}^{kd} \frac{\sqrt{l^2 - (kw)^2}}{l}\exp(-\alpha l)dl, \quad \mathfrak{I}_2 = \frac{2}{h\pi k}\int_{kh}^{kd} \frac{\sqrt{l^2 - (kh)^2}}{l}\exp(-\alpha l)dl$$

## S4. Fraction of light emitted to the waveguiding mode

The emitted light from a fluorophore (quantum dot, organic dye, etc.) in a polymer/glass slab will experience total internal reflection for angles at the air interface larger than a critical angle $\alpha_c$. In the most common case for a glass or a polymer: $n = 1.5$, $n_{air} = 1$ and the critical angle $\alpha_c$:

$$\sin(\alpha_c) = \frac{1}{n}$$

i.e. $\alpha_c \approx 42°$. Thus, for the emitter in a rectangular slab with six facets there are six cones with the angle $2\alpha_c$, where the emitted light can escape. Solid angle of the cone (surface of a spherical cap) for a unity radius sphere is

$$S_1 = 2\pi(1 - \cos\alpha_c)$$

So the fraction of the emitted light through one facet is:

$$\delta_1 = \frac{S_1}{S} = \frac{2\pi(1 - \cos\alpha_c)}{4\pi} = \frac{\left(1 - \frac{\sqrt{n^2 - 1}}{n}\right)}{2}$$

Considering only emitted light through top and bottom facets as losses the total useful fraction of the emission is then ($n = 1.5$):

$$\delta = 1 - 2\delta_1 = \frac{\sqrt{n^2 - 1}}{n} \approx 75\%$$

## S5. Effect of scattering by fluorophores

If the total loss is governed by the scattering instead (absorption-free matrix and re-absorption free fluorophore) then the optical efficiency can be also evaluated from the optical path length distribution. Let the linear scattering coefficient be $\alpha_{sc}[1/cm]$. Probability for the photon to travel optical path $l'$ before reaching the edge is $q(l')dl'$. Probability of not being scattered within distance $l'$ is $\exp(-\alpha_{sc}l')$. These photons will contribute to the total optical efficiency similarly to the absorption case above:

$$\chi_0(\alpha_{sc}) = \int_0^{lmax} q(l') \cdot \exp(-\alpha_{sc}l') \, dl' = f(\alpha_{sc})$$

In addition, there will be photons, which underwent scattering into the waveguiding mode. Probability of being scattered within distance $l'$ is $1 - \exp(-\alpha_{sc}l')$. If $\delta$ is a fraction of waveguided light after a scattering event ($\delta$=75% for n=1.5) the probability to reach the edge after one scattering event is:

$$\chi_1(\alpha_{sc}) = \delta \cdot \int_0^{lmax} q(l') \cdot (1 - \exp(-\alpha_{sc}l'))dl' \cdot \int_0^{lmax} q(l') \cdot \exp(-\alpha_{sc}l') \, dl'$$

$$\chi_1(\alpha_{sc}) = \delta \cdot (1 - f(\alpha_{sc})) \cdot f(\alpha_{sc})$$

A Markov process is considered, where there is no memory in the system. The total probability for a photon to reach the edge becomes then a geometrical series (sum of probabilities for no scattering, one scattering, two scattering events, etc.):

$$\chi(\alpha_{sc}) = \sum_{i=0}^{\infty} \chi_i = f(\alpha_{sc})\left[1 + \delta \cdot (1 - f(\alpha_{sc})) + \left(\delta \cdot (1 - f(\alpha_{sc}))\right)^2 + \ldots\right] = \frac{f(\alpha_{sc})}{1 - \delta \cdot (1 - f(\alpha_{sc}))}$$

## S6. Effect of several loss mechanisms present simultaneously

Now consider two processes taking place simultaneously: scattering and matrix absorption. First, photons experiencing no scattering and no absorption will contribute to the total signal:

$$\psi_0(\alpha_{sc}, \alpha) = \int_0^{lmax} q(l') \cdot \exp(-\alpha_{sc}l') \cdot \exp(-\alpha l') \, dl' = f(\alpha_{sc} + \alpha)$$

Then photons after one scattering event and without subsequent scattering and absorption. While every scattering event sets back to zero the travelled distance for scattering, the optical path for absorption continues. So the exact history of scattering becomes important. To take into account this fact one can introduce a probability *density* to scatter at a point $l'$ (in the absence of other processes):

$$p_{sc}(l') = \alpha_{sc} \cdot \exp(-\alpha_{sc} l')$$

which is a properly normalized probability density function. Then in the system where scattering and absorption coexist the probability density to scatter at a point $l'$ without being absorbed before is:

$$p_{sc}(l') \int_{l'}^{\infty} p_{ab}(x) dx$$

where a similar notation of the probability density $p_{ab}$ is introduced for the pure absorption process.

Additional conditions of no subsequent scattering and absorption can be added as:

$$p_{sc}(l') \int_{l'}^{\infty} p_{ab}(x) dx \cdot \delta \cdot \exp(-\alpha_{sc} l_2) \cdot \exp(-\alpha l_2)$$

where $l_2$ is a photon path taken to reach the device edge after the scattering event. If $l'$ varies in between $(0; l_1)$ the integrated *probability* becomes:

$$\delta \exp(-(\alpha_{sc} + \alpha) l_2) \int_0^{l_1} \alpha_{sc} \exp(-\alpha_{sc} l') \exp(-\alpha l') \, dl'$$
$$= \exp(-(\alpha_{sc} + \alpha) l_2) \frac{\delta \alpha_{sc}}{\alpha_{sc} + \alpha} [1 - \exp(-(\alpha_{sc} + \alpha) l_1)]$$

Finally taking into account probability to have photon path $l_1$ as $q(l_1) dl_1$ and $l_2$ as $q(l_2) dl_2$ (again Markov process without memory in the system considered) one obtains after integration from zero to $l_{max}$ for both path stretches $l_{1,2}$ the input from the photons experienced one scattering event:

$$\psi_1(\alpha_{sc}, \alpha) = f(\alpha_{sc} + \alpha) \cdot \delta \alpha_{sc} \frac{[1 - f(\alpha_{sc} + \alpha)]}{\alpha_{sc} + \alpha}$$

Continuing in the same manner for two scattering events without subsequent scattering and absorption:

$$\psi_2(\alpha_{sc}, \alpha) = f(\alpha_{sc} + \alpha) \cdot (\delta \alpha_{sc})^2 \left( \frac{[1 - f(\alpha_{sc} + \alpha)]}{\alpha_{sc} + \alpha} \right)^2$$

So the resulting probability can be again represented through geometrical series:

$$\psi(\alpha_{sc}, \alpha) = \sum_{i=0}^{\infty} \psi_i(\alpha_{sc}, \alpha) = \frac{f(\alpha_{sc} + \alpha)}{1 - \delta \cdot \frac{\alpha_{sc}}{\alpha_{sc} + \alpha} \cdot [1 - f(\alpha_{sc} + \alpha)]}$$

This formula turns into the expression for scattering only scenario for a non-absorbing matrix ($\alpha = 0$).

A similar result can be derived for the case of re-absorption instead of scattering:

$$\phi(\alpha_{re}, \alpha) = \frac{f(\alpha_{re} + \alpha)}{1 - \delta \cdot QY \cdot \frac{\alpha_{re}}{\alpha_{re} + \alpha} \cdot [1 - f(\alpha_{re} + \alpha)]}$$

When re-absorption and scattering both exist in the system one can show in a similar manner as above:

$$\varphi(\alpha_{sc}, \alpha_{re}) = \frac{f(\alpha_{sc} + \alpha_{re})}{1 - \frac{\delta \cdot \alpha_{sc} + \delta \cdot QY \cdot \alpha_{re}}{\alpha_{sc} + \alpha_{re}}(1 - f(\alpha_{sc} + \alpha_{re}))}$$

A general solution for the optical efficiency, following derivations above, is:

$$g(\alpha_{sc}, \alpha_{re}, \alpha) = \frac{f(\alpha_{sc} + \alpha_{re} + \alpha)}{1 - \frac{\delta \cdot \alpha_{sc} + \delta \cdot QY \cdot \alpha_{re}}{\alpha_{sc} + \alpha_{re} + \alpha}(1 - f(\alpha_{sc} + \alpha_{re} + \alpha))} \qquad (2)$$